\documentclass{article}

\usepackage{arxiv}

\usepackage[utf8]{inputenc} 
\usepackage[T1]{fontenc}    
\usepackage{hyperref}       
\usepackage{url}            
\usepackage{booktabs}       
\usepackage{amsfonts}       
\usepackage{nicefrac}       
\usepackage{microtype}      
\usepackage{lipsum}

\usepackage{setspace}
\usepackage{caption}
\usepackage{subcaption}

\usepackage[figuresright]{rotating}

\title{Warfarin dose estimation on multiple datasets with automated hyperparameter optimisation and a novel software framework}

\author{
  Gianluca Truda \\
  Department of Computer Science\\
  University of Cape Town\\
  Cape Town, South Africa \\
  \texttt{trdgia001@myuct.ac.za} \\
   \And
 Patrick Marais \\
  Department of Computer Science\\
  University of Cape Town\\
  Cape Town, South Africa \\
  \texttt{patrick@cs.uct.ac.za} \\
}

\begin{document}
\maketitle

\begin{abstract}
Warfarin is an effective preventative treatment for arterial and venous thromboembolism, but requires individualised dosing due to its narrow therapeutic range and high individual variation. Many machine learning techniques have been demonstrated in this domain. This study evaluated the accuracy of the most promising algorithms on the International Warfarin Pharmacogenetics Consortium dataset and a novel clinical dataset of South African patients. Support vectors and linear regression were amongst the top performers in both datasets and performed comparably to recent stacked ensemble approaches, whilst neural networks were one of the worst performers in both datasets. We also introduced genetic programming to automatically optimise model architectures and hyperparameters without human guidance. Remarkably, the generated models were found to match the performance of the best models hand-crafted by human experts. Finally, we present a novel software framework (\href{https://pypi.org/project/warfit-learn/}{Warfit-learn}) for warfarin dosing research. It leverages the most successful techniques in preprocessing, imputation, and parallel evaluation, with the goal of accelerating research and making results in this domain more reproducible. 
\end{abstract}


\section{Introduction} \label{sec:intro}
Many individuals suffer from blood clots that lead to arterial and venous thromboembolism. The standard method for treating these conditions is the use of anticoagulant drugs such as vitamin K antagonists, the most widely used of which is warfarin. Whilst effective, the drug has a narrow therapeutic range and severe side-effects at extreme concentrations. This makes the precise dosing of warfarin an important concern for clinicians. Unfortunately, warfarin metabolism differs across individuals based on age, weight, genetics, diet, drug interactions, and various pre-existing conditions \cite{Jonas2009,Wells1994}. 
To standardise the process of anticoagulant monitoring, the World Health Organisation adopted the international normalised ratio (INR) \cite{kirkwood1983}. INR has become the standard measurement for anticoagulation monitoring around the world \cite{Poller2004}. 

Many studies \cite{Liu2015,Liu2012a,Sharabiani2015,Hu2012,Zhou2014,Grossi2014,Sharabiani2013, Ma2018} have looked at applying statistical models to the problem of individualised warfarin dosing. Accurate models improve the ability of clinicians to prescribe the correct warfarin doses to their patients, whilst minimising the time required to do so. They also reduce the risk of severe haemorrhaging in patients and the number of visits required to establish a therapeutic dose \cite{Poller2008, McDonald2008, Kim2010}. Unfortunately, warfarin datasets are small and noisy, requiring the use of specialised data transformations and highly-optimised learning algorithms. Breakthrough techniques are of significance to the medical research community and could lead to improvements in decision support for those administering warfarin therapy.

The learning algorithm chosen to produce the dosing model has a notable impact on accuracy and robustness. Moreover, the precise hyperparameters chosen for the algorithm can drastically affect performance. 

Unfortunately, replicating the precise methodology of previous studies has proven difficult, as the methods for preprocessing, imputation, data stratification, and evaluation vary. Moreover, these methods are typically described textually, with open-sourced software implementations being unavailable prior to this study. We introduce a new software framework (Warfit-learn) that implements the best available techniques. We validate the framework by using it to compare a variety of established and novel algorithms on two different warfarin datasets. 

\subsection{Related work}

\subsubsection{International warfarin pharmacogenetics consortium (IWPC)} 
The seminal work in warfarin dose estimation with pharmacogenetic data made use of 5052 patient records from the initial version of the dataset utilised by later studies \cite{Klein2009}. Researchers split the dataset randomly with an 80/20 ratio into derivation and validation sets. The study selected the two key metrics that would come to be used in almost all future work  --  mean absolute error (MAE) and percentage of patients within 20\% of the therapeutic dose (PW20). They found that a multiple linear regression technique yielded better results than the existing clinical algorithms, with an MAE of $8.5 \pm 1.7$ mg/week in the validation cohort \cite{Klein2009}.

\subsubsection{Liu et al. 2015}
A notable study is that of Liu et al. in 2015 \cite{Liu2015}, which compared the average performance of 9 learning algorithms on the IWPC dataset \cite{Whirl2012}. They removed patients missing height, weight, age, or genotype data, and patients not at a stable warfarin dose  --  leaving 4798 patients remaining. Using libraries in the R language, they implemented 9 algorithms. They obtained the average performance of each algorithm in terms of PW20 and MAE with 100 rounds of 80/20 re-sampling from the filtered dataset. The focus of their study was evaluating a range of off-the-shelf algorithms across dosage ranges and racial groups, but their top results (PW20 = 46.35\%, MAE = 8.84) in the combined cohort set an initial benchmark for performance on the IWPC data. This result was achieved with a multivariate adaptive regression splines algorithm, though the authors note that Bayesian additive regression trees and support vector machines were also high-performing algorithms.

\subsubsection{Ma et al. 2018}
Another notable study is that of Ma et al. in late 2018 \cite{Ma2018}, which made advances using ensemble techniques and improved imputation of missing data. Utilising the same IWPC dataset \cite{Whirl2012}, Ma et al. imputed missing weight and height values using a linear regression model. For imputing height, the variables were weight, race, and sex. For imputing weight, they were height, race, and sex. They imputed missing values for the \textit{VKORC1 rs9923231} genotype using the IWPC's formula \cite{IWPCsupplementary} based on race and linkage disequilibrium in \textit{VKORC1}. Even after excluding outliers, 5743 subjects remained in the data set, resulting in a cohort nearly 20\% greater than that of Liu et al. The raw warfarin dose was square rooted to compensate for the skewed distribution of the variable. The researchers made use of a stacked generalisation (stacking) framework to implement heterogeneous ensembles. Results were obtained through 100 rounds of 80/20 re-sampling. Their findings suggest that stacked generalisation ensembles are significantly more accurate than the existing isolated models on the IWPC dataset. Crucially, however, these high-performing stacked ensembles were given extra input parameters that were not accessible to the other algorithms. In addition to the 11 common parameters used in the IWPC and Liu et al. models \cite{Whirl2012,Liu2015}, their stacked ensembles included indicators for diabetes mellitus, heart failure, valve replacement, use of statins, and additional \textit{VKORC1} genotypes. Despite acknowledging the use of additional parameters in their methodology, the authors still claim to have performed a direct comparison. Although they appear to have successfully utilised these additional parameters, the inconsistency of the methodology renders their findings inconclusive. Considering that hyperparameter tuning was performed on the entire dataset, it is hard to differentiate their reported accuracy improvements from possible overfitting effects. With our open-sourced Warfit-learn framework, we hope to reduce the possibility of such inconsistencies in future work.

\subsection{Model optimisation}
The most common approach to model development is for an expert to repeatedly train models with various learning algorithms and tweak their hyperparameters until performance is maximised. Ensemble methods -- such as bagging, boosting, and stacking -- combine the outputs from several models to obtain better overall performance on unseen data. This has been found to increase model performance in a number of warfarin dosing studies \cite{Liu2015,Cosgun2011a,Hu2012,Ma2018}. Another approach is the use of automatic machine learning (autoML), which employs meta-algorithms to automate the task of optimisation. 

In the past, autoML approaches have focused on optimising subsets of the machine learning pipeline \cite{hutter2015}. Grid search is a common form of hyperparameter optimisation based on a brute-force search of algorithm hyperparameters, with the drawback of computational complexity. Conversely, random evaluation within the search space typically discovers an effective hyperparameter set more quickly than exhaustive search \cite{bergstra2012}. A trade-off between the expensive (but comprehensive) exhaustive search and the efficient (but uncertain) random evaluation is desired. A well-known approach to this trade-off is evolutionary algorithms. Indeed, the use of genetic programming  --  a subset of evolutionary algorithms  --  has been found to produce better models than human experts in a number of tasks \cite{zutty2015, Olson2016, hornby2011, fredericks2013, forrest2009, spector2008}. This promising avenue to autoML was explored in our paper.

\subsection{Genetic programming}
Genetic programming emulates the process of natural selection as an optimisation strategy. In the context of this study, the smallest components are machine learning algorithms and data processors. Each is encoded as a gene, with parameters changing according to defined mutation and crossover rates. As with biological evolution, successful genes propagate through the population and the most successful combinations of those genes seed the next generation \cite{banzhaf1998}. This increases performance over time. By simulating many generations with well-chosen evolutionary hyperparameters and population sizes, novel meta-algorithms with high accuracy can be produced. Genetic programming has been shown to develop effective systems in a number of mathematical and computational domains \cite{hornby2011,fredericks2013,forrest2009,spector2008}. 

This study made use of the open-sourced Tree-based Pipeline Optimisation Tool (TPOT) to generate high-performing models through genetic programming \cite{Olson2016}. The TPOT framework was developed atop the Distributed Evolutionary Algorithms in Python framework \cite{fortin2012}. Each individual in the population is a tree-based machine learning pipeline. Each node in the tree is a pipeline operator, which can be (1) a preprocessor, (2) a decomposer, (3) a feature selector, or (4) a supervised learning algorithm. All of these operators are based on Scikit-learn \cite{scikit-learn} implementations. These pipelines handle the data from feature extraction through to hyperparameter optimisation. The pipeline operators (and their parameters) are encoded as a gene sequence, which evolves through the genetic programming process \cite{Olson2016}.

To prevent a tendency toward overly-complex pipelines, TPOT makes use of Pareto optimisation to optimise two objectives simultaneously  --  maximising pipeline fitness, whilst minimising pipeline complexity (measured as the number of operators). Over many generations of evolution, TPOT's implementation of the genetic programming algorithm evolves the pipelines by adding new pipeline operators that improve fitness, and by removing redundant or detrimental pipeline operators. The fittest overall pipeline is saved as the representative pipeline at the end of a TPOT run \cite{Olson2016}. In other domains, TPOT has been shown to produce a significant improvement over basic machine learning methods, with little involvement from users \cite{Olson2016}. This study utilised TPOT to automatically evolve optimal architectures and hyperparameters for warfarin dose estimation. The resulting estimators performed on par with those manually optimised by machine learning experts.

\section{Materials and methods} \label{sec:methods}
\begin{figure}
    \centering
    \includegraphics[width=12cm]{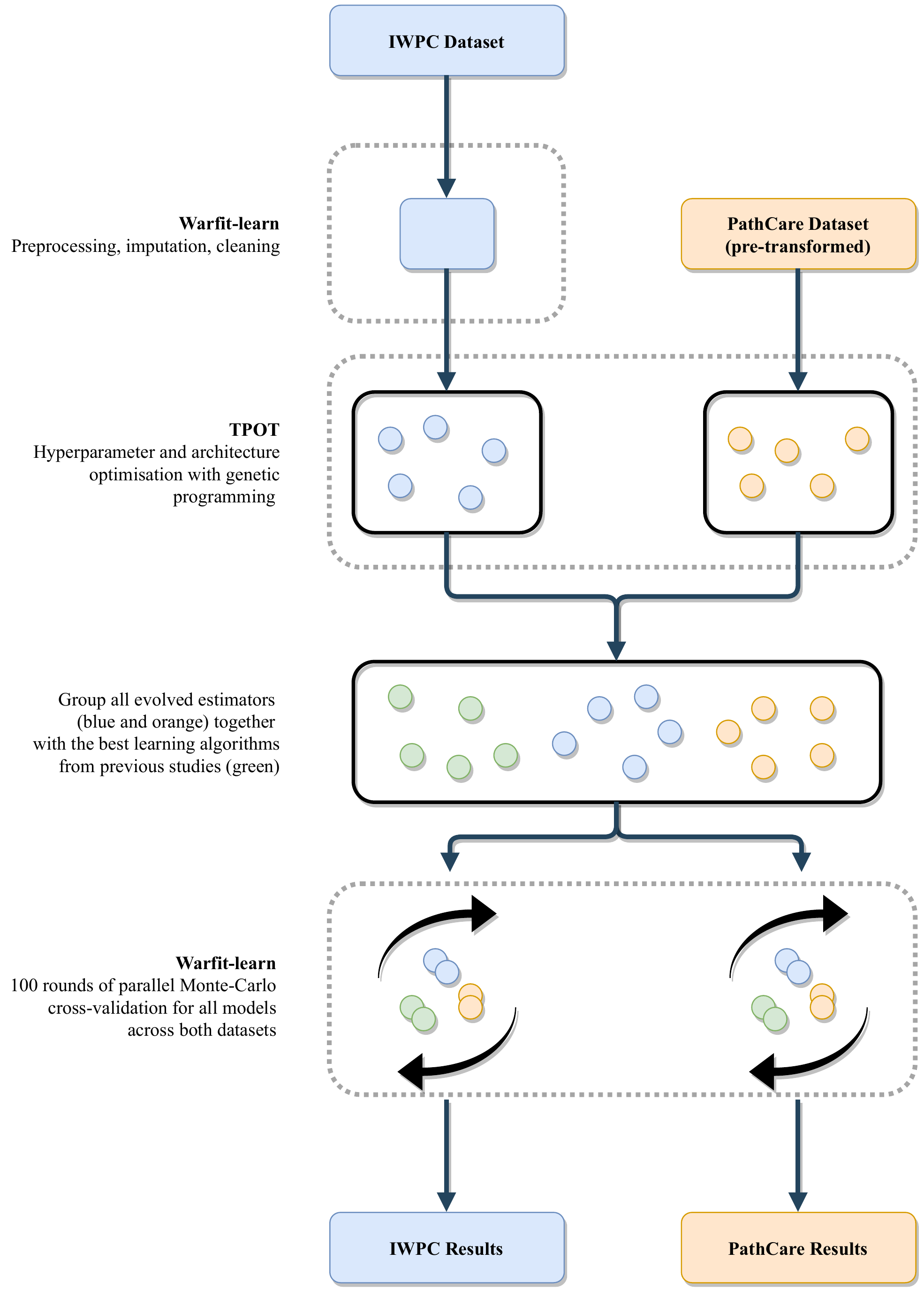}
    \caption{An illustration of the methodology applied in this study, highlighting where our Warfit-learn software framework was utilised and how a collection of estimators were evaluated across multiple datasets.}
    \label{fig:methodology}
\end{figure}

An illustration of the study's methodology can be found in Figure \ref{fig:methodology} and highlights where our Warfit-learn software framework was applied.

\subsection{Datasets}
Two datasets of warfarin records were used for this study. The standard IWPC dataset was used primarily for comparing new techniques to those in existing literature, whilst a novel dataset from South Africa was used to evaluate how well estimators generalised to a different clinical context and parameter set. Both datasets have similar distributions of weekly warfarin dose, INR, and age (see Table \ref{tab:cohort_breakdown}).

The International Warfarin Pharmacogenetics Consortium (IWPC) dataset \cite{Whirl2012} of 6256 patients has been used in a number of notable studies \cite{Liu2015, Sharabiani2015, Klein2009, Ma2018} and is the standard reference point for new approaches to automated warfarin dosing. The dataset was compiled collaboratively and includes data from 22 research groups from 9 countries \cite{Klein2009}. The dataset is publicly available at \href{https://www.pharmgkb.org/downloads}{pharmgkb.org/downloads}.

The novel dataset was provided by PathCare, a private pathology group in South Africa. Unlike the IWPC dataset, no pharmacogenetic data is available and very limited clinical data is present. A total of 4621 patients from the dataset achieved their target INR range and were considered in this study (see Table \ref{tab:cohort_breakdown}). As with the IWPC dataset, patients were de-identified prior to distribution. To guarantee the confidentiality of the data, only the authors had access to it during the study. Ethical approval was obtained from the Science Research Ethics Committee at the University of Cape Town and the project was sanctioned by the PathCare Research Committee. To protect the patients, the data is not publicly available.

\begin{table}
\centering
\caption{Analysis of the characteristics of the International Warfarin Pharmacogenetics Consortium (IWPC) subjects and subjects from a novel South African dataset (PathCare) included in this study.}

\begin{tabular}{lllll}

\textbf{} & \multicolumn{2}{c}{\textbf{IWPC}} & \multicolumn{2}{c}{\textbf{PathCare}} \\ \hline
Size & \multicolumn{2}{c}{5741} & \multicolumn{2}{c}{4621} \\ \hline
\multicolumn{5}{l}{\textbf{Warfarin dose (mg/week)}} \\ \hline
Mean & \multicolumn{2}{l}{31.97} & \multicolumn{2}{l}{29.63} \\
Median & \multicolumn{2}{l}{28.0} & \multicolumn{2}{l}{27.50} \\
Range & \multicolumn{2}{l}{2.1 – 230.0} & \multicolumn{2}{l}{0.0 – 210.0} \\
Standard deviation & \multicolumn{2}{l}{16.77} & \multicolumn{2}{l}{16.04} \\ \hline
\multicolumn{5}{l}{\textbf{Therapeutic INR}} \\ \hline
Mean & \multicolumn{2}{l}{2.15} & \multicolumn{2}{l}{2.46} \\
Median & \multicolumn{2}{l}{2.34} & \multicolumn{2}{l}{2.40} \\
Range & \multicolumn{2}{l}{0.0 – 6.1} & \multicolumn{2}{l}{1.8 – 3.9} \\
Standard deviation & \multicolumn{2}{l}{0.82} & \multicolumn{2}{l}{0.33} \\ \hline
\textbf{Age} & \textbf{Count} & \textbf{\%} & \textbf{Count} & \textbf{\%} \\ \hline
10 - 19 & 15 & 0.3 & 2 & 0.0 \\
20 - 29 & 130 & 2.3 & 31 & 0.7 \\
30 - 39 & 249 & 4.3 & 106 & 2.3 \\
40 - 49 & 580 & 10.1 & 276 & 6.0 \\
50 - 59 & 1133 & 19.7 & 490 & 10.6 \\
60 - 69 & 1401 & 24.4 & 942 & 20.4 \\
70 - 79 & 1539 & 26.8 & 1143 & 24.7 \\
80 - 89 & 660 & 11.5 & 1127 & 24.4 \\
90+ & 34 & 0.6 & 204 & 4.4 \\ \hline
\textbf{Sex} & \textbf{Count} & \textbf{\%} & \textbf{Count} & \textbf{\%} \\ \hline
Females & 2466 & 43.0 & 2443 & 52.9 \\
Males & 3275 & 57.0 & 2178 & 47.1 \\
Missing & 0 & 0 & 0 & 0 \\ \hline

\end{tabular}
\begin{flushleft}
Note: These are the cohorts after the preprocessing and imputation stages.
\end{flushleft}
\label{tab:cohort_breakdown}
\end{table}

\begin{table}
\centering
\caption{Continued: Analysis of the characteristics of the International Warfarin Pharmacogenetics Consortium (IWPC) subjects and subjects from a novel South African dataset (PathCare) included in this study.}

\begin{tabular}{lllll}

\textbf{} & \multicolumn{2}{c}{\textbf{IWPC}} & \multicolumn{2}{c}{\textbf{PathCare}} \\ \hline

\multicolumn{5}{l}{\textbf{Height (cm)}} \\ \hline
Mean & \multicolumn{2}{l}{167.98} & \multicolumn{2}{l}{-} \\
Median & \multicolumn{2}{l}{167.64} & \multicolumn{2}{l}{-} \\
Range & \multicolumn{2}{l}{125 – 202} & \multicolumn{2}{l}{-} \\
Standard deviation & \multicolumn{2}{l}{10.58} & \multicolumn{2}{l}{-} \\ \hline
\multicolumn{5}{l}{\textbf{Weight (kg)}} \\ \hline
Mean & \multicolumn{2}{l}{78.79} & \multicolumn{2}{l}{-} \\
Median & \multicolumn{2}{l}{76.0} & \multicolumn{2}{l}{-} \\
Range & \multicolumn{2}{l}{30 – 238} & \multicolumn{2}{l}{-} \\
Standard deviation & \multicolumn{2}{l}{22.32} & \multicolumn{2}{l}{-} \\
\hline
\textbf{Race (OMB)} & \textbf{Count} & \textbf{\%} & \textbf{Count} & \textbf{\%} \\ \hline
White & 3095 & 53.9 & - & - \\
Asian & 1515 & 26.4 & - & - \\
Black & 665 & 11.6 & - & - \\
Missing / mixed race & 466 & 8.1 & - & - \\ \hline
\textbf{VKORC1 Genotype (Imputed)} & \textbf{Count} & \textbf{\%} & \textbf{Count} & \textbf{\%} \\ \hline
A/A & 1695 & 29.5 & - & - \\
A/G & 2058 & 35.8 & - & - \\
G/G & 1884 & 32.8 & - & - \\
Unknown & 104 & 1.8 & - & - \\ \hline
\textbf{CYP2C9 Genotype} & \textbf{Count} & \textbf{\%} & \textbf{Count} & \textbf{\%} \\ \hline
*1/*1 & 4230 & 73.7 & - & - \\
*1/*2 & 755 & 13.2 & - & - \\
*1/*3 & 482 & 8.4 & - & - \\
*1/*5 & 0 & 0.0 & - & - \\
*1/*6 & 0 & 0.0 & - & - \\
*2/*2 & 58 & 1.0 & - & - \\
*2/*3 & 68 & 1.2 & - & - \\
*3/*3 & 20 & 0.3 & - & - \\
Unknown & 128 & 2.2 & - & - \\ \hline
\textbf{Indications} & \textbf{Count} & \textbf{\%} & \textbf{Count} & \textbf{\%} \\ \hline
Enzyme inducer* & 61 & 1.1 & - & - \\
Amiodarone & 280 & 4.9 & 131 & 2.8 \\
Smoker & 482 & 8.4 & - & - \\
\hline
\end{tabular}
\begin{flushleft}
Notes: These are the cohorts after the preprocessing and imputation stages. The OMB racial categories were used in the IWPC dataset and are defined by the 1997 Office of Management and Budget (OMB) standards for the U.S. Census Bureau. * = carbamazepine, phenytoin, rifampin, rifampicin.
\end{flushleft}
\label{tab:cohort_breakdown2}
\end{table}

\newpage

\subsection{Data cleaning}
For the IWPC dataset, this study replicated the work of Ma et al. \cite{Ma2018} for filtering the patient records into a viable cohort. This was done to verify their approach and to provide a consistent methodology for performance comparisons. The process is detailed in their paper and in the source code of this study. In this domain, missing data has typically been handled by either dropping whole records \cite{Liu2015} or imputing missing values \cite{Ma2018,Sharabiani2013,Klein2009}. Because the IWPC data is limited in size, dropping missing values leads to significant data loss and is detrimental to performance. This study leveraged the most successful imputation techniques from previous work -- missing values for \textit{VKORC1 rs9923231} were imputed using the IWPC's formula \cite{IWPCsupplementary} based on race and linkage disequilibrium, whilst missing height and weight were imputed with a regression model as per the work of Ma et al. \cite{Ma2018}. This process is built into the Warfit-learn preprocessing module.

\subsection{Parameter selection}
For the purposes of direct comparison to baseline, this study used the same parameter set defined by the IWPC \cite{Whirl2012} and utilised by Liu et al. \cite{Liu2015}.  For the PathCare dataset, three parameter sets were evaluated using standard regression techniques. The top-performing parameter set was selected and used for further evaluation. This set included the sex and age of the patients; as well as records of their aspirin, paracetamol, and amiodarone use; and their history of atrial fibrillation, deep vein thrombosis, and heart-valve replacement. All categorical parameters were vectorised into sparse format so that the resulting feature set was purely numerical. 

\subsection{Preprocessing}
The input features were scaled prior to training using the Scikit-learn implementation of \texttt{StandardScaler}  --  which scales the data to unit variance. This desensitised the algorithms to the magnitudes of features. In both datasets, the weekly dose (in mg) was transformed to its square root in order to correct for the skewed distributions. This was then set as the target feature during training. During evaluation, the model predictions were squared and compared with the original values -- maintaining clinical relevance. Warfit-learn provides the option to perform this rooting and squaring automatically.

\subsection{Learning algorithms used} \label{ssec:learning_algs_used}
A basic linear regression model -- similar to that of the IWPC \cite{Klein2009} -- was required as a baseline for model performance. This was compared against many of the best-performing algorithms described by Liu et al. \cite{Liu2015} and Ma et al. \cite{Ma2018}. All algorithms were implemented using the Scikit-learn library \cite{scikit-learn} in Python 3, with the use of MLxtend \cite{raschka_2018} to implement stacked ensembles. The implementation details and key hyperparameters can be found in Table \ref{tab:learning_algs}. Additionally, a naïve algorithm was implemented to establish a lower bound on what could be considered as legitimate modelling. This algorithm simply learns the median of the training data and then predicts that value for all subsequent cases. 

\begin{sidewaystable}[] 
\centering
\caption{Learning algorithms replicating the work of the IWPC \cite{Klein2009}, Liu et al. \cite{Liu2015}, and Ma et al. \cite{Ma2018} that were compared in this study.}
\begin{tabular}{lll}
\textbf{Name} & \textbf{Implementation} & \textbf{Key hyperparameters} \\ \hline
LR & sklearn.linear\_model.LinearRegression & normalize=False, fit\_intercept=True \\
SVR & sklearn.svm.LinearSVR & epsilon=0.0, tol=0.0001, C=1.0, loss='epsilon\_insensitive' \\
SV & sklearn.svm.SVR & kernel='linear', cache\_size=1000 \\
RR & sklearn.linear\_model.Ridge & alpha=1.0 \\
BRT & sklearn.ensemble.GradientBoostingRegressor & loss='least squares', learning\_rate=0.1, n\_estimators=100 \\
GBT & sklearn.ensemble.GradientBoostingRegressor & learning\_rate=0.1, loss='lad', max\_depth=4 \\
NN & sklearn.neural\_network.MLPRegressor & hidden\_layer\_sizes=(100, ), activation='logistic', solver='lbfgs' \\
Stacked\_SV & mlxtend.regressor.StackingCVRegressor & regressors=\{GBT, SV, NN\},  meta\_regressor=SV, cv=5 \\
Stacked\_RR & mlxtend.regressor.StackingCVRegressor & regressors=\{GBT, RR, NN\}, meta\_regressor=RR, cv=5 \\ \hline
\end{tabular}
\begin{flushleft}
The estimators used in the stacked ensembles had the same hyperparameters as the standalone estimators of the same name, as per the methodology of Ma et al. \cite{Ma2018}. LR = (multiple) linear regression, SV(R) = support vector (regression), RR = ridge regression, BRT = boosted regression trees, GBT = gradient-boosted trees,  NN = neural network. Note the slight implementation differences between SV (following the Ma et al. implementation) and SVR (following a generic implementation more similar to that of Liu et al.)
\end{flushleft}
\label{tab:learning_algs}
\end{sidewaystable}

\subsection{Clinical and statistical metrics}
Appropriate metrics were chosen to assess both the clinical and statistical accuracy of the models. An important clinical consideration is that INR has a therapeutic range of $\pm 0.5$ in most patients \cite{Albers2001} and using tighter target ranges for maintenance dosing does not achieve any therapeutic advantage \cite{Meier2007}. The chosen metrics accounted for that, and were consistent with metrics used in related studies  --  allowing direct results comparisons. Moreover, the use of multiple metrics for performance (one clinically-relevant, the other statistically-relevant) is a worthwhile methodology to employ as it prevents the kind of systemic overfitting that occurs in other domains where a single metric is dominant.
\begin{enumerate}
\item \textbf{Mean absolute error (MAE)} is used widely across warfarin estimation studies \cite{Liu2015,Liu2012a,Sharabiani2015,Tan2012,Hu2012,Zhou2014,Grossi2014,Sharabiani2013, Ma2018}.
\begin{equation}
MAE = \frac{\sum_{i=1}^{n}|y_i - \hat{y}_i|}{n}
\end{equation}
where $y_i$ is the actual dose and $\hat{y}_i$ is the predicted dose.
\item \textbf{Percentage of patients with dose estimates within 20\% of the actual therapeutic dose (PW20)} was used by a number of notable studies \cite{Klein2009,Grossi2014,Zhou2014,Liu2015, Ma2018}. It reflects the fact that being within 0.5 points of the target INR is clinically sufficient \cite{Meier2007}.
\begin{equation}
PW20 = \frac{\sum_{i=1}^{n}f(p_i)}{n} \times 100\%
\end{equation}
where $f(p_i)$ for patient $p_i$ is 1 if $0.8y_i < \hat{y}_i < 1.2y_i$, else 0; $\hat{y}_i$ is the predicted dose, and $y_i$ is the true therapeutic dose.
\end{enumerate}

\subsection{Evaluation protocol}
Monte Carlo Cross-Validation (MCCV) was performed using the standard \texttt{train\textunderscore test\textunderscore split} function in Scikit-learn. This performs resampling \textit{without} replacement. All models were trained and evaluated using 100 iterations of MCCV for each filtered dataset. In each iteration, the data was randomly split into a training set and a testing set in an 80/20 ratio. The model was then fit on the training set and evaluated in terms of PW20 and MAE on the testing set. The means of these results and 95\% confidence intervals were calculated for each model on each dataset based on the percentile method. 

\subsection{Warfit-learn software library}
\label{ssec:warfit_learn}
Extending past studies in warfarin dose prediction required the development of a host of data cleaning and preprocessing tools, evaluation routines, and scoring functions. To remove the need for repeating this effort in future work, this study implements a new software framework for Python-based development of warfarin dose estimation models. The aim of this library is to allow researchers to more easily replicate and extend the work of their colleagues in this domain. This can help to reduce confusion about methodology and enhance reproducibility. The library focuses on four areas, namely:
\begin{enumerate}
\item Seamless dataset loading, cleaning, and preprocessing.
\item Standardised implementations of scoring functions.
\item Multithreaded model training and evaluation using standardised resampling techniques.
\item Full interoperability with the Python scientific stack \cite{scipy2011}, Pandas \cite{pandas2010}, and Scikit-learn \cite{scikit-learn}.
\end{enumerate}

This makes it possible to replicate studies on the IWPC data with only a beginner's level of Python experience. Supervised learning models can be constructed using the popular Scikit-learn implementations, which are handled natively by the framework. Libraries that extend the Scikit suite  --  such as MLxtend \cite{raschka_2018}  --  can be used directly with minimal configuration. Parallel processing on a user-defined number of CPU cores is used to achieve speedup on the resampling performed during evaluation and is wrapped in a user-friendly function. Functions for metrics like percentage of patients within 20\% of therapeutic dose (PW20) are implemented and automatically called during evaluation. Final results are produced as Pandas dataframes, which can be exported to CSV files or LaTeX tables with a single command. The entire framework is available as open source software at \url{http://pypi.org/project/warfit-learn} under a GNU GPLv3 license. All results for this study were obtained using the above tools in version $0.2$ of Warfit-learn.

\subsection{Optimisation with genetic programming}
TPOT \cite{Olson2016} was used to generate high-performing machine learning pipelines through genetic programming. In each run, a preprocessed version of the dataset was given as input to TPOT and many generations of computation yielded the best performers. TPOT accepts bespoke scoring functions as its evolutionary fitness function. The functions for PW20 and MAE were trialled independently as fitness functions. We also explored a combination of these metrics. Attempting to both maximise PW20 and minimise MAE simultaneously would have transformed the task into a multi-objective optimisation. A popular approach to overcome these complexities is \textit{scalarisation}, which numerically combines both objectives into a single optimisation function \cite{eiben2015introduction}. However, this requires a top-down definition of appropriate weightings for both the PW20 and MAE. To avoid introducing additional hyperparameters, we defined a new hybrid function as the ratio of the maximised (PW20) to minimised (MAE) objectives: 
\begin{equation}
\mathrm{hybrid} = \frac{\mathrm{PW20}}{\mathrm{MAE}^2}    
\end{equation}Because PW20 was encoded as a percentage, typical values were in the range $[20, 50]$, whilst MAE was usually below 15. We thus squared the MAE denominator so that the two values would be more similar in range, whilst also boosting the effects of variations in the error component. 

Many instances of TPOT were run on a multi-core machine using different evolutionary hyperparameters. The number of generations ranged from 50 to 1000, the number of offspring from 5 to 100, and the \textit{k} used in \textit{k}-fold cross-validation from 5 to 30. Some runs used the raw weekly dose as the target feature and others used the square root of weekly dose. Over all the instances run on both datasets, the best performers (according to final validation scores within TPOT) were evaluated against the human-optimised algorithms described in other studies.

\newpage
\section{Results and discussion} \label{sec:results}
\begin{figure}[h!]
     \centering
     \begin{subfigure}[b]{0.49\textwidth}
         \centering
         \includegraphics[width=\textwidth]{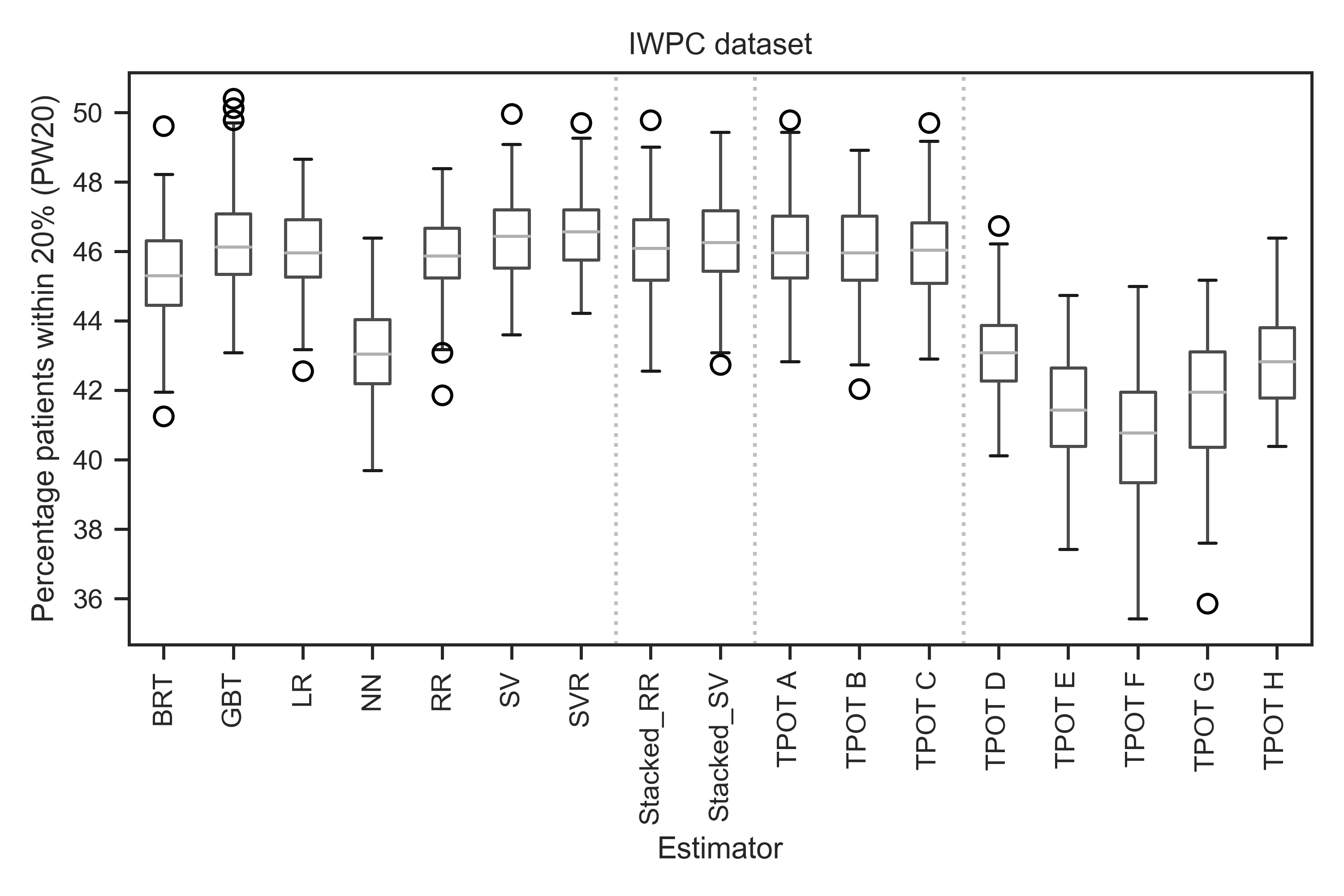}
     \end{subfigure}
     \hfill
     \begin{subfigure}[b]{0.49\textwidth}
         \centering
         \includegraphics[width=\textwidth]{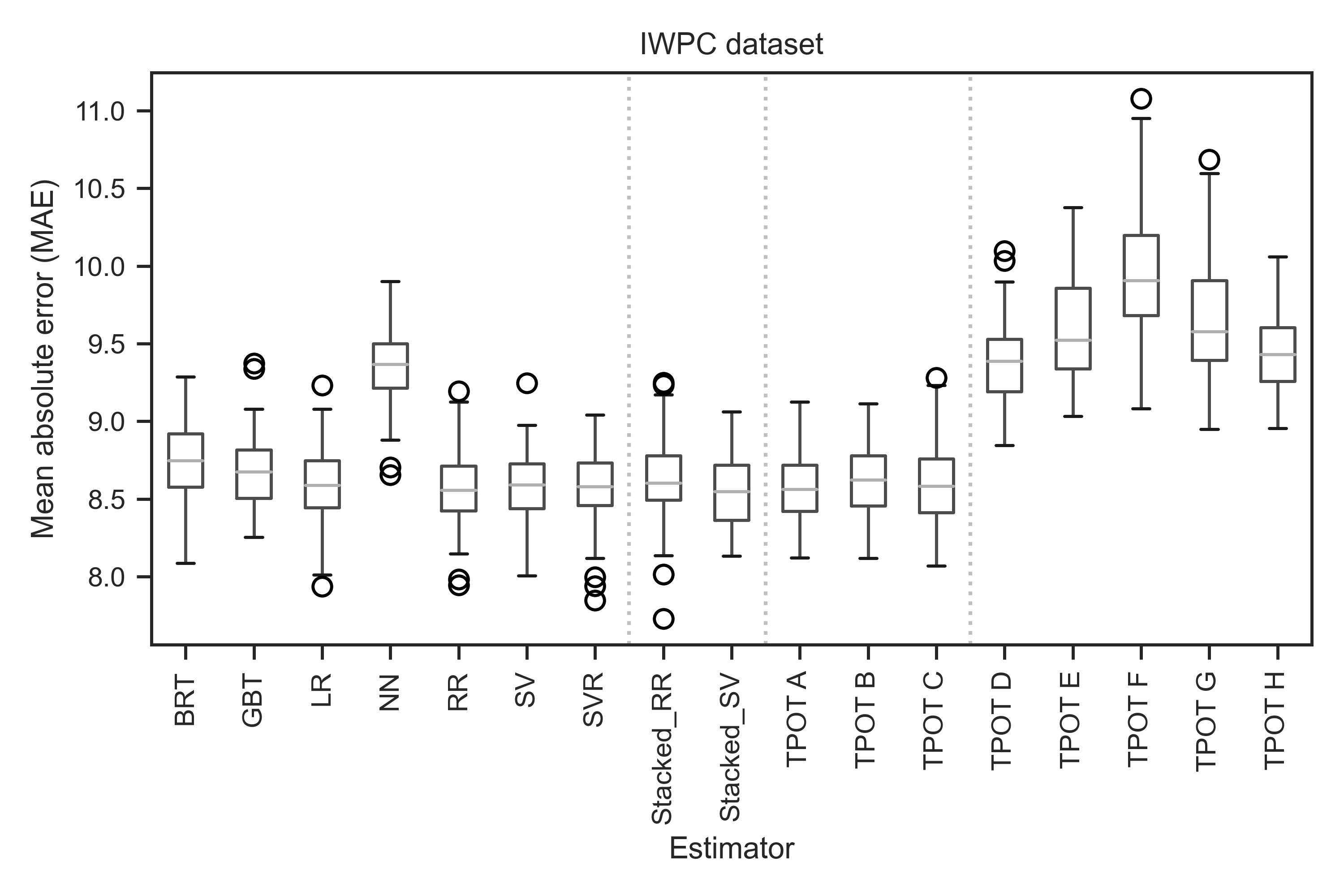}
     \end{subfigure}
     \hfill
     \begin{subfigure}[b]{0.49\textwidth}
         \centering
         \includegraphics[width=\textwidth]{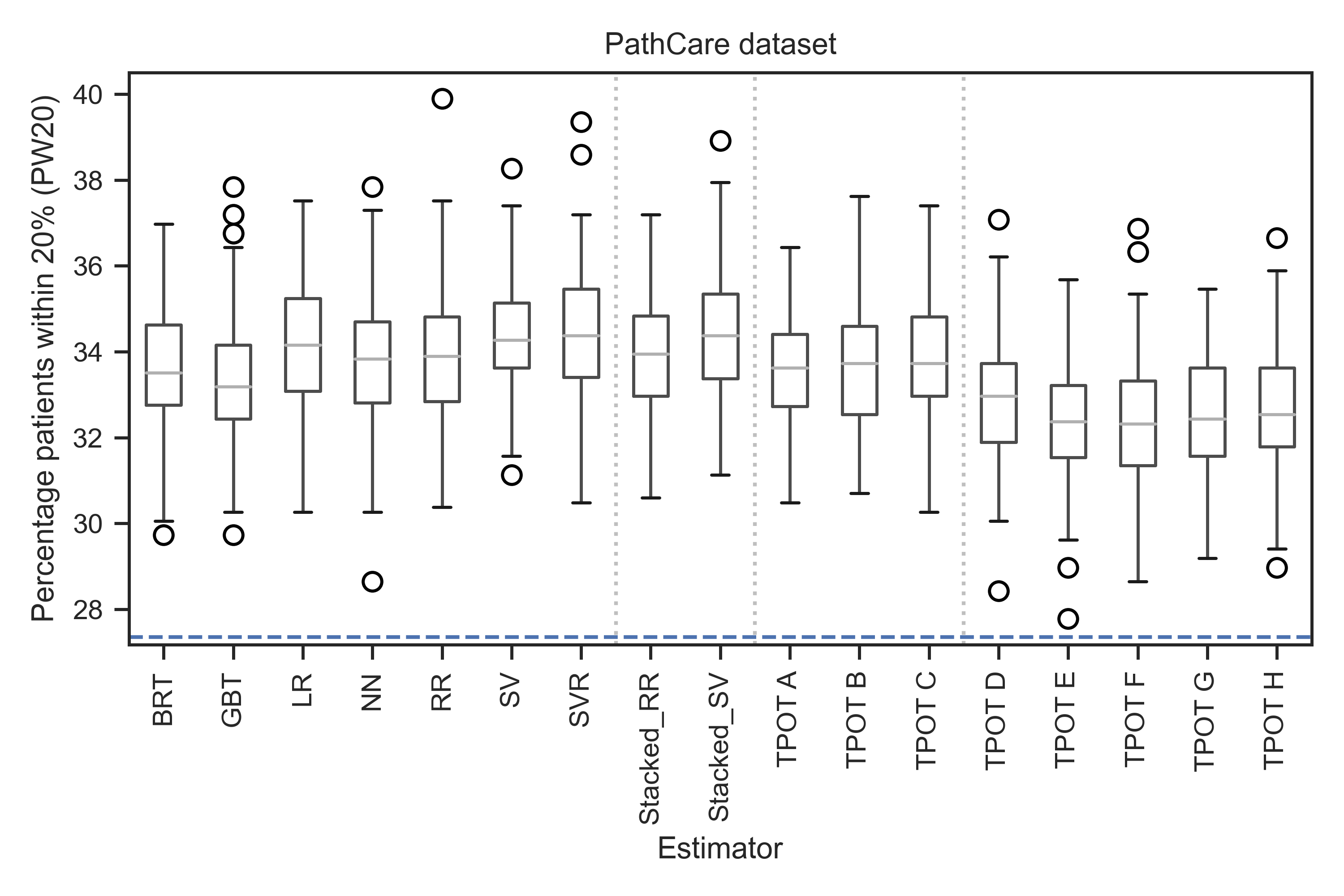}
     \end{subfigure}
     \hfill
     \begin{subfigure}[b]{0.49\textwidth}
         \centering
         \includegraphics[width=\textwidth]{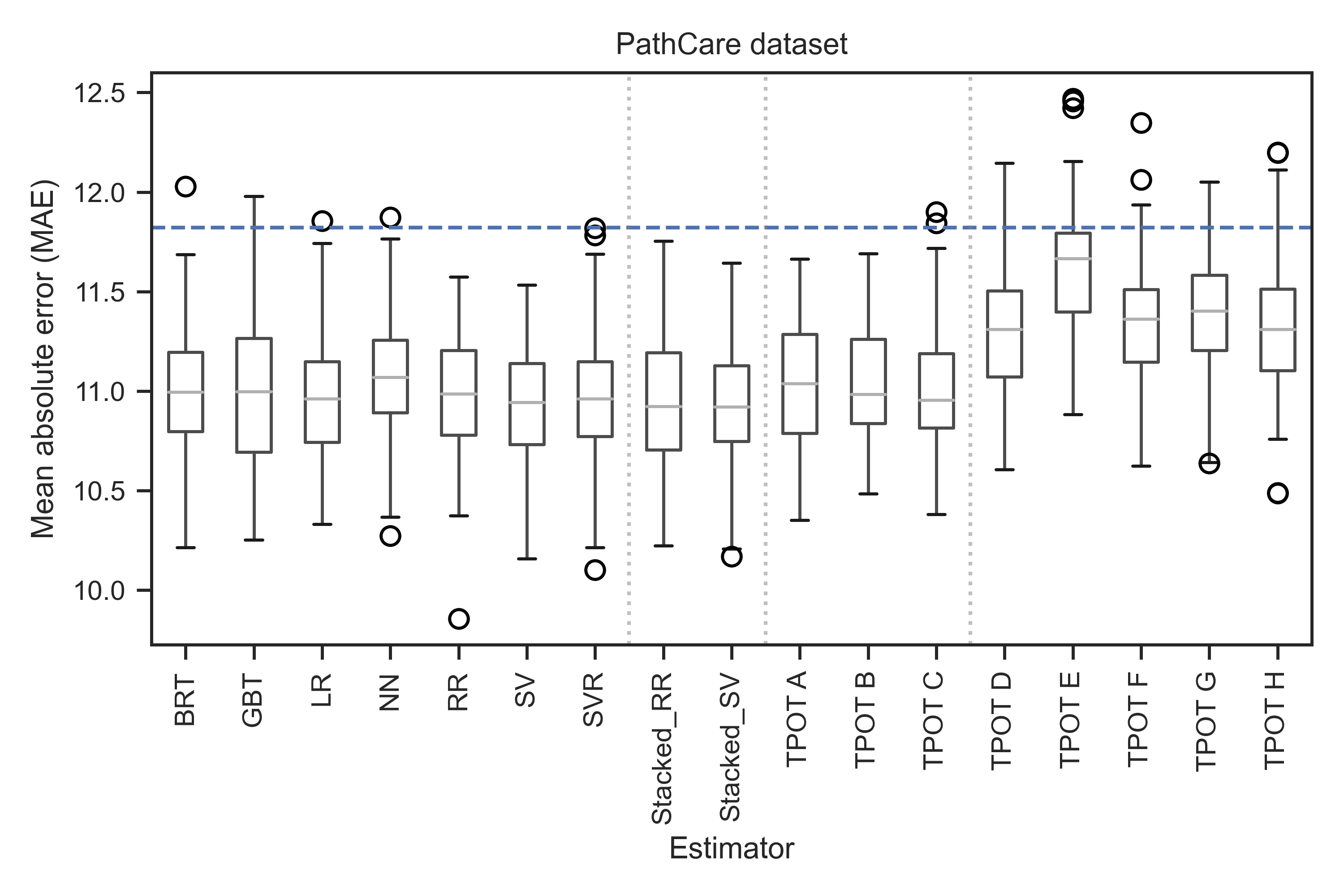}
     \end{subfigure}
     \hfill
   
    \caption{Box plots showing the distributions of all estimators after 100 rounds of MCCV evaluation on both the IWPC (top row) and PathCare (bottom row) datasets. Each box marks the first and third quartiles with a line at the median. The whiskers extend to the full range of the data. Results outside 1.5 times the interquartile range are considered outliers and marked with circles. The dotted vertical lines delineate the traditional algorithms, the stacked ensembles from Ma et al. \cite{Ma2018}, the IWPC-optimised TPOT architectures, and the PathCare-optimised architectures. The dashed blue lines on the PathCare plots (bottom row) indicate the mean performance of a naïve estimator. The axes are automatically scaled to best illustrate the variation between estimators. BRT = boosted regression trees, GBT = gradient-boosted trees, LR = (multiple) linear regression, NN = neural network, RR = ridge regression, SV(R) = support vector (regression). The naïve estimator always predicted the median of the training data and was used as a lower bound on what is considered genuine modelling. See the implementation details for each estimator in Table \ref{tab:learning_algs}. Details of the pipelines generated by TPOT can be found with the published source code for this project.}
    \label{fig:results_boxplots}
\end{figure}

The performance of all estimators is compared across datasets in Table \ref{tab:results_general}. For easy comparison with previous studies, the mean and 95\% confidence intervals are reported for both the MAE and PW20 metrics. In order to more clearly illustrate and compare the distributions of the results, box plots are presented in Figure \ref{fig:results_boxplots}.

\begin{sidewaystable}
\centering
\caption{Results of all estimators after 100 rounds of MCCV evaluation on both the IWPC and novel (PathCare) datasets, with the best performers in each group emphasised.}
\begin{tabular}{lllll}
          \ & 
          \multicolumn{2}{c}{\textbf{IWPC Dataset}} & \multicolumn{2}{c}{\textbf{PathCare Dataset}} \\
          \hline
     \textbf{Estimator} &                 \textbf{PW20 (95\% CI)} &                \textbf{MAE (95\% CI)} &                 \textbf{PW20 (95\% CI)} &                  \textbf{MAE (95\% CI)} \\
\hline
           BRT &  45.32 (42.47–47.70) &   8.75 (8.35–9.18) &  33.61 (31.19–36.27) &  10.99 (10.40–11.52) \\
           GBT &  46.31 (43.73–49.74) &   8.67 (8.32–9.07) &  33.35 (30.42–36.60) &  10.99 (10.37–11.60) \\
            LR &  45.99 (43.38–48.22) &   8.59 (8.11–9.05) &  34.08 (30.81–36.82) &  10.98 (10.39–11.67) \\
            NN &  43.12 (40.69–45.87) &   9.36 (8.90–9.80) &  33.79 (31.19–37.14) &  11.07 (10.56–11.67) \\
            RR &  45.96 (43.33–48.30) &   8.57 (8.16–9.03) &  33.99 (31.19–36.92) &  10.98 (10.45–11.53) \\
            SV &  46.41 (44.07–48.74) &   8.58 (8.12–8.96) &  34.40 (31.88–36.81) &  10.93 (10.36–11.45) \\
           \textbf{SVR} &  \textbf{46.56 (44.60–49.13)} &  \textbf{8.58 (8.05–9.00)} &  \textbf{34.46 (31.77–37.14)} &  \textbf{10.95 (10.33–11.60)} \\
\hline
    Stacked\_RR &  46.01 (43.29–48.30) &   8.63 (8.19–9.14) &  33.83 (30.86–36.55) &  10.95 (10.39–11.54) \\
    \textbf{Stacked\_SV} &  \textbf{46.32 (43.81–48.48)} &   \textbf{8.55 (8.21–8.99)} &  \textbf{34.45 (31.78–37.62)} &  \textbf{10.93 (10.26–11.57)} \\
    
    \hline
    
        TPOT A* &  46.03 (43.03–48.26) &   8.61 (8.17–9.06) &  33.65 (31.29–36.54) &  11.05 (10.58–11.66) \\
        TPOT B* &  46.03 (43.56–48.83) &   8.59 (8.16–9.06) &  33.91 (31.83–36.49) &  11.01 (10.44–11.69) \\
         \textbf{TPOT C*} &  \textbf{46.08 (43.73–48.53)} &   \textbf{8.56 (8.14–9.02)} &  \textbf{33.53 (31.03–35.85)} &  \textbf{11.03 (10.49–11.53}) \\
    TPOT D\textsuperscript{\dag} &  43.09 (40.55–45.88) &   9.38 (8.93–9.88) &  32.86 (30.32–35.75) &  11.29 (10.75–11.98) \\
 TPOT E\textsuperscript{\dag} &  41.44 (38.19–44.30) &  9.59 (9.07–10.17) &  32.32 (29.62–34.99) &  11.63 (11.03–12.29) \\
 TPOT F\textsuperscript{\dag} &  40.54 (36.38–43.78) &  9.96 (9.28–10.86) &  32.37 (29.95–35.20) &  11.35 (10.74–11.93) \\
 TPOT G\textsuperscript{\ddag} &  41.75 (38.68–44.92) &  9.64 (9.05–10.58) &  32.53 (30.21–35.25) &  11.39 (10.83–11.95) \\
 TPOT H\textsuperscript{\ddag} &  42.82 (40.64–45.60) &   9.43 (8.99–9.89) &  32.67 (29.83–35.41) &  11.33 (10.80–12.07) \\
 \hline
 Naïve & 34.10 (33.85–34.35) &  12.45 (12.39–12.50) &  27.35 (27.10–27.61) &  11.82 (11.75–11.89) \\
 \hline
\end{tabular}
\begin{flushleft}
MAE = mean absolute error. PW20 = percentage of patients within 20\% of therapeutic dose. Values are provided as means with 95\% confidence intervals. All estimators utilised the same parameter set for the IWPC data, as defined by the seminal IWPC study \cite{Klein2009}. * = Evolved on IWPC dataset. \dag = Evolved on PathCare dataset using raw weekly dose. \ddag = Evolved on PathCare dataset using square root of weekly dose. BRT = boosted regression trees, GBT = gradient-boosted trees, LR = (multiple) linear regression, NN = neural network, RR = ridge regression, SV(R) = support vector (regression). The naïve estimator always predicted the median of the training data and was used as a lower bound on what is considered genuine modelling. See the implementation details for each estimator in Table \ref{tab:learning_algs}. Details of the pipelines generated by TPOT can be found with the published source code for this project.
\end{flushleft}
\label{tab:results_general}
\end{sidewaystable}

\newpage
\subsection{Validation of methodology}
The methodology of this study was validated by referencing the results of previous work on warfarin dose estimation. The original IWPC study reported an MAE of 8.5 mg/week (95\% CI: 8.1–8.6) for their linear regression (LR) model \cite{Klein2009}. However, they processed the cohort differently.

The study by Ma et al. evaluated models on an identical dataset to our study, reporting an MAE of 8.53 (95\% CI: 8.08–8.99) for their LR model \cite{Ma2018}. By comparison, the MAE for the LR model in this study was 8.59 mg/week (95\% CI: 8.11–9.05). This indicates that our preprocessing and evaluation methodology were valid. It also justifies the original IWPC finding that, when adequate preprocessing is performed, linear regression produces outstanding results on the IWPC dataset \cite{Klein2009}. This, coupled with its easy interpretability, makes LR hard to rule out as the default warfarin estimation model. However, as datasets grow in size and variety, LR can fail to model some relationships that more complex approaches succeed in capturing. It also tends to be less robust to outliers than ensemble approaches \cite{bishop2006pattern}, which could challenge its clinical applicability. 

The shapes of the performance distributions in Figure \ref{fig:results_boxplots} are quite symmetric and consistent with normal distributions, which affirms the use of confidence intervals based on the percentile method.  Despite some large differences in average performance over the 100 resamplings, the distributions still overlap considerably, highlighting how much performance can vary depending on the specific patients in the training and testing sets. With few exceptions, the ranges of PW20 and MAE scores for even the most consistent algorithms were considerably greater than the differences between the mean scores, suggesting that both datasets have a high degree of inter-patient variation that is unaccounted for in the available parameters. 

The purpose of mirroring our methodology on the PathCare dataset was to evaluate how well the estimators generalise to a different clinical context and parameter set. The relationships of the IWPC results are certainly reflected in the PathCare results, both in terms of relative scores (Table \ref{tab:results_general}) and variance (Figure \ref{fig:results_boxplots}). However, they are much less pronounced. 

To verify that effective modelling took place, we made use of a naïve estimator that simply predicted the median of its training set for each resampling. On the IWPC dataset, it had a PW20 of 34.10 (95\% CI: 33.85–34.35) and an MAE of 12.45 (95\% CI: 12.39–12.50), making it the lowest performer by a considerable amount. On the PathCare dataset, it achieved a PW20 of 27.35 (95\% CI: 27.10–27.61) and an MAE of 11.82 (95\% CI: 11.75–11.89). The other estimators only marginally outperformed the naïve baseline on the PathCare dataset in terms of MAE, but well outperformed it in terms of PW20 -- the clinically-relevant metric. This affirms that even a limited number of clinical parameters are sufficient for the estimators to learn some relationships from the data and make meaningful predictions. For the PathCare dataset, the mean performance of the naïve baseline is overlaid in the box plots of Figure \ref{fig:results_boxplots} for visual reference.

\subsection{Performance of traditional techniques}

Support vector regression (SV and SVR) was a top-performing algorithm, whilst the neural network (NN) was one of the worst performing algorithms. The relative performance of the NN and SVR models was similar to that reported by Liu et al. \cite{Liu2015}. Their NN (MAE 9.82, PW20 41.27) was by far their worst performer, whilst their SVR (MAE 8.96, PW20 45.88) was one of their best performers. Similarly, this study's NN (MAE 9.36, PW20 43.12) was the worst-performing standalone algorithm by a considerable margin, whilst our SVR (MAE 8.58, PW20 46.56) was the best. 

Unlike most statistical models, neural networks are considered to be good at learning feature representations automatically, which is especially useful when there are many features. The IWPC dataset has very few parameters compared to typical deep learning datasets. Moreover, these parameters are heavily-engineered during the preprocessing stage, which likely minimises any effects of the representation learning. It is also widely held that NN performance scales with the size of the dataset and the number of hidden layers \cite{Choromanska2014}. Because the IWPC dataset contains very few records (by deep learning standards) using more than one hidden layer has little-to-no effect. These factors may explain why the NN consistently failed to perform on both datasets.

Our results replicate the findings of Ma et al. \cite{Ma2018} for LR, SV, and ridge regression (RR) algorithms, with very similar confidence intervals. But, despite replicating the implementations outlined in their methodology, our results for NN, gradient boosting trees (GBT), Stacked SV, and Stacked RR were notably dissimilar. In the case of the ensemble estimators, this was likely due to their decision to use a much larger parameter set on their stacked implementations than they fed to other models. In contrast, our study made use of the same parameter set across all estimators (as standardised by notable previous works \cite{Klein2009, Liu2015}). Another factor may have been the difference in GBT implementations (Scikit-learn vs. LightGBM) between studies. This discrepancy once again affirms the need for a standardised set of tools for replicating and evaluating warfarin dosing models. 

Despite the difference in absolute scores between studies, the stacked ensembles were still among the top performers on the IWPC data, with the Stacked SV in particular achieving an MAE of 8.55 and a PW20 of 46.32. This supports the use of the stacked generalisation approach to improve warfarin dose estimation. However, with a much smaller effect size than Ma et al. reported, it may be difficult to justify the use of these techniques over the more-explainable linear regression model in clinical deployment. 

Despite the absence of pharmacogenetic data and many clinical parameters in the PathCare dataset, the traditional algorithms performed reasonably well. LR (MAE 10.98, PW20 34.08) and SV (MAE 10.93, PW20 34.40) topped the list, along with the Stacked SV ensemble (MAE 10.93, PW20 34.45). However, there was very little difference in performance on the PathCare dataset amongst the traditional algorithms. The range of the distributions were also much wider on the PathCare dataset than the IWPC dataset (see Figure \ref{fig:results_boxplots}). The fact that all the estimators still outperformed the naïve baseline indicates that there were indeed predictive relationships to be found in the PathCare dataset, but it seems that the precise choice of algorithm was of much less importance with so few parameters available. Algorithms with higher bias, like LR and SV, performed marginally better, whereas algorithms with lower bias, like the ensembles, appear to have generalised less well. 

\subsection{Performance of evolved architectures}
\begin{figure}[!h]
    \centering
    \includegraphics[width=8cm]{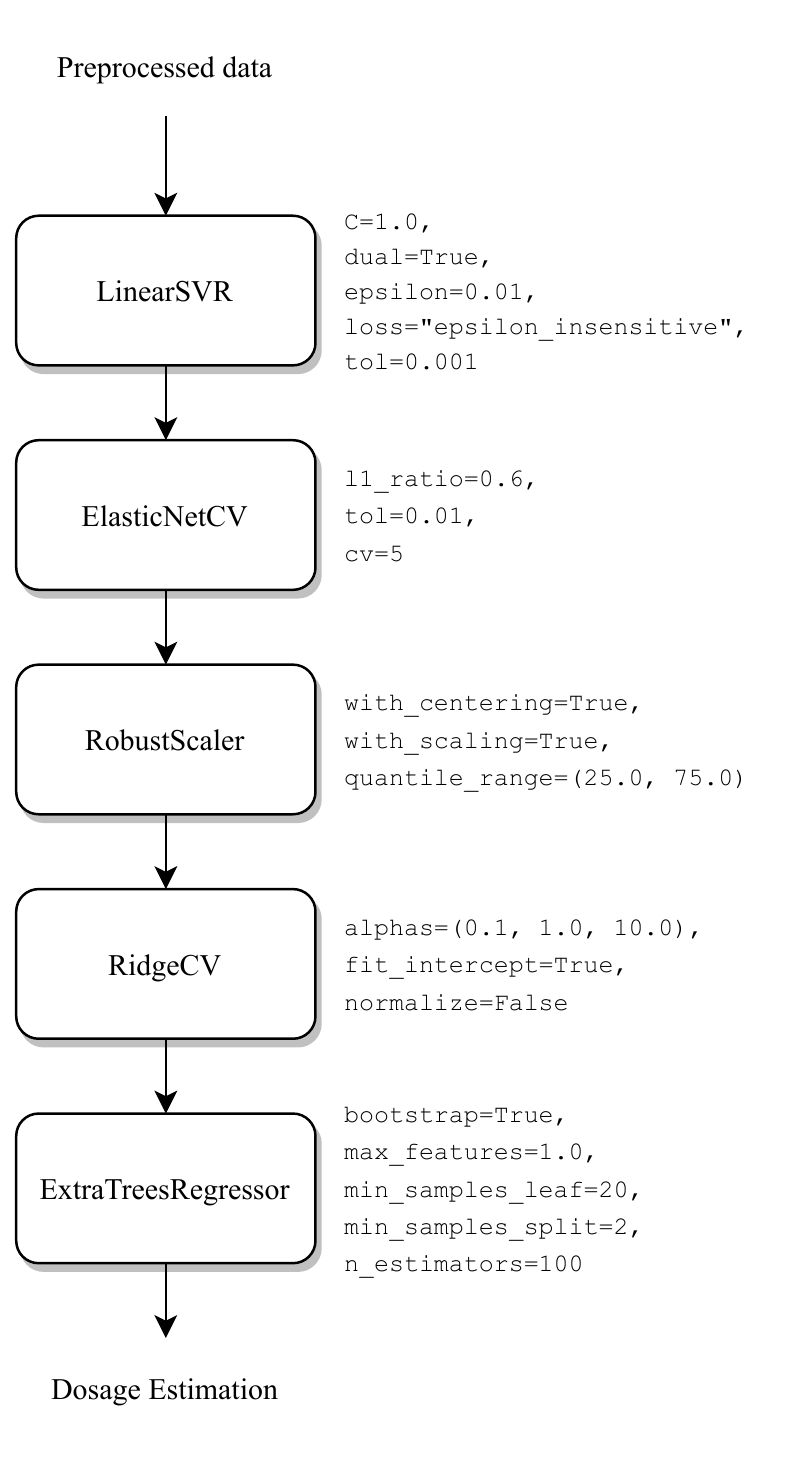}
    \caption{Illustration of the highest-performing architecture and hyperparameters evolved on the IWPC dataset using TPOT. The Python implementation of this architecture can be found along with the project \href{https://github.com/gianlucatruda/warfit-learn}{source code}.}
    \label{fig:best_tpot}
\end{figure}

The architectures developed using a genetic programming approach (TPOT) had extremely varied results depending on which dataset they were evolved on. TPOT C was an estimator pipeline evolved on the IWPC dataset that produced the best results of the evolved models (MAE 8.56, PW20 46.08). It was only marginally outperformed by the SV and Stacked SV approaches (with highly similar CIs), but outperformed the Stacked RR estimator on the IWPC dataset. The structure and hyperparameters of the estimator are illustrated in Figure \ref{fig:best_tpot}.

Architectures evolved on the IWPC dataset (A, B, and C) performed well on the IWPC dataset, with almost identical distributions. This is remarkable when considering that they were generated with no human tuning. It is also notable that three different evolutionary histories converged to have such consistent performance. This result concurs with previous findings \cite{zutty2015, Olson2016} that autoML can attain results comparable to those of models engineered by experts with domain knowledge.

Generally, the TPOT architectures that performed well on the IWPC data also performed well on the PathCare data. This could indicate that the IWPC dataset captures sufficient information about the parameter-target relationships to generalise, but may also be a result of limitations in the PathCare dataset. 

\newpage
The architectures evolved on the PathCare dataset (D through H) performed significantly worse than those evolved on the IWPC dataset (A through C) regardless of which dataset they were trained and evaluated on. Moreover, the distributions of results were much wider, indicating low generalisation. These performance differences are best illustrated by the jarring shift in distributions seen in the top row of Figure \ref{fig:results_boxplots}. These results suggest that the PathCare dataset has a lower ratio of signal to noise compared to the IWPC dataset and support prior claims \cite{Pirmohamed2013,Sconce2005,Wadelius2007} that pharmacogenetic parameters yield more generalised models.

All but one of the architectures evolved with TPOT utilised stacking at some point in their pipelines. This may reflect a general tendency within the TPOT implementation but, when considered with the high performance of Ma et al.'s stacked ensembles, may indicate the robustness of stacking for warfarin dose estimation. Only TPOT C (the best overall) and TPOT E included SVR in their pipelines. This is strange considering the high level of performance seen with SVR across both datasets. TPOTs B and C (high performers on the IWPC dataset) made use of trees (with boosting) and/or lasso regression. TPOTs D-H (evolved on PathCare) all relied on a KNN regressor somewhere in their pipeline, but performed poorly across datasets.

\subsection{Limitations} 
The IWPC dataset was compiled by 22 research groups, each with different protocols and equipment. This resulted in noisy data and some missing values, which lower the predictive accuracy of models. It is also known that the impact of \textit{CYP2C9} and \textit{VKORC1} genes varies across races \cite{Liu2015}. Current pharmacogenetic data is mostly derived from White and Asian OMB racial groups in developed nations (where genetic testing is available), so it is likely that current pharmacogenetic implementations impart bias upon models.

The PathCare dataset did not include clinical information such as height, weight, and race  --  all of which have been shown to improve dosing accuracy. It would, therefore, be imprudent to extrapolate very much from the performance differences of models on this dataset. Nevertheless, it served as a useful validation that top-performing algorithms can generalise to novel clinical contexts with different parameter sets. Because of these different parameters, it was not possible to directly train models on one dataset and evaluate their performance on another. It was, however, possible to evolve the architectures and hyperparameters on one dataset and then train and evaluate models on the other dataset -- which was done and is presented in Table \ref{tab:results_general} and Figure \ref{fig:results_boxplots}.

There are also technical limitations in the performance of the genetic programming tools. As Olson et al. point out themselves, TPOT becomes quite slow as the size of a dataset increases. Even on the tiny (by machine learning standards) IWPC dataset, a single run could take days to complete. There are plans to integrate other autoML libraries and heuristics to seed the populations with strong pipelines and reduce the number of computations to convergence \cite{Olson2016}. Additionally, the use of highly-parallel hardware, such as graphics processing units (GPUs), offers enormous potential for speedup in execution. Because TPOT is built atop the popular, open-source Scikit-learn framework, it benefits from modularity and standardisation. Both the \textit{H2O4GPU} project and Nvidia's \textit{cuML} suite mirror the Scikit-learn interface, but move computations to the GPU. At the time of writing, there is at least one ongoing project to bring this functionality into TPOT. 

Unlike linear regression, the other models do not produce human-interpretable formulae for clinical warfarin dosing. However, trained models could be used to run inference on real patient data and produce dosage recommendations to help guide clinicians. Further work is needed to produce robust models that can be deployed as decision-support systems in a clinical context.

\section{Conclusions} \label{sec:conclusion}
This study presented the Warfit-learn software framework, which implements the best published techniques for warfarin dosage modelling. The framework was validated by replicating the results of several prior studies on the same IWPC dataset. Warfit-learn's parallel evaluation methodology was utilised to compare the performance of several estimators across both the IWPC dataset and a novel dataset of South African patients provided by PathCare. The estimators included traditional learning algorithms that have shown promise in warfarin dosing, two stacked generalisation ensembles described in previous work, and a collection of learning pipelines evolved using a genetic programming approach to architecture and hyperparameter optimisation. 

As in previous studies, linear regression was shown to be a reliable technique. Support vector regression was the best performing traditional algorithm, whilst neural networks performed poorly. Recent ensemble approaches -- namely stacked generalisation -- were shown to be effective across datasets, but the effect size was not as large as previously reported.

This study found that genetic programming was effective at evolving robust architectures and hyperparameters for warfarin dosage modelling. All three of the estimators evolved on the IWPC dataset were amongst the top-performing models, performing on par with the best expert-designed ensembles. Inspection of the architectures revealed a preference for stacked ensembles. If these results are not unique to warfarin dosing, they suggest that automated machine learning may be a promising method for the future of software-assisted dosing. These findings should be explored on other dosage estimation problems in future work. Other advances in this domain can be made by compiling comprehensive warfarin dosing datasets from more diverse cohorts, and using them to train robust models for safe clinical deployment.

In addition to the performance comparisons and evolutionary optimisation techniques, this paper presented a much-needed software framework for evaluating new techniques on the IWPC dataset. Warfit-learn eliminates the need for future researchers to re-create tools, and maintains consistency by providing a standardised testbed. It is our hope that Warfit-learn will make future studies significantly easier and more reproducible.

\bibliographystyle{unsrt}  
\bibliography{mybibfile}

\end{document}